\documentstyle[aps,prl]{revtex}
\begin{document}
\draft \twocolumn[\hsize\textwidth\columnwidth\hsize\csname
@twocolumnfalse\endcsname 
\title{ On the nature of the transition from the spontaneously dimerized
 to the N\'{e}el phase  in  
 the two-dimensional
$J_{1}-J_{2}$ model }
\author{Valeri N. Kotov$^1$ and Oleg P. Sushkov$^2$
 } \address{$^1$ Department of Physics,
University of Florida, Gainesville, Florida 32611-8440
\\ $^2$School of Physics, University of New South
 Wales, Sydney 2052, Australia}

\maketitle
\begin{abstract}
 We analyze the spectrum of the 2D S=1/2 frustrated Heisenberg model
 near the transition from the spontaneously dimerized spin-liquid
 phase into the N\'{e}el ordered phase. Two excitation branches: 
the triplet magnon, and the collective singlet mode, both become gapless 
 at the transition point. However we find that the 
 length scales associated with these modes are
  well separated at the quantum transition.
While in the quantum disordered phase the singlet excitation has finite spectral weight and reflects the existence of
 spontaneous dimer order, near the transition point 
 the size of the
 singlet bound state grows exponentially with the correlation length,
and hence the quasiparticle residue is exponentially small.
Therefore the critical dynamics remains in the $O(3)$ universality class
in spite of the four gapless modes.

\end{abstract}

 ]


 Quantum phase transitions between magnetically ordered and  
 disordered phases   can take place at $T=0$ by varying the exchange interactions
 which can drive the spin-spin correlations from long-range behavior, characterized
 by an infinite correlation length, towards a short-range regime, typical for disordered
 phases. An example of a quantum model which exhibits such transitions 
 is the two-dimensional (2D), $S=1/2$ Heisenberg
 antiferromagnet (HAFM) on a square lattice. While for uniform nearest-neighbor interactions
 the HAFM has long-range N\'{e}el order in the ground state with sub-lattice magnetization
 $M\approx 0.3$ \cite{mano}, inclusion of additional interactions, such as dimerization
 and/or
 frustration, leads to increased quantum fluctuations and ultimately vanishing
 of $M$ at a critical coupling. Examples of transitions caused by local alternation
 of the exchange couplings are
the dimerized HAFM \cite{rajiv,dimer}, the two-layer HAFM \cite{chub,us} and
 the $CaV_4O_9$ lattice (1/5-th depleted square lattice) \cite{cavo}. In these cases
 the local dimer or plaquette correlations eventually win over the long-range
 N\'{e}el order, leading to a non-magnetic ground state. 
  Another route towards a magnetically disordered ground state  is  introduction
 of frustrating second-neighbor interactions ($J_2$), in addition to the nearest-neighbor
 ones ($J_{1}$) (see Fig.1.). The  N\'{e}el order disappears at $(J_2/J_1)_{c} \approx 0.4$
 in this case \cite{linear,igarashi,exact1,ising}.

 An important issue concerning the quantum transitions mentioned above
 is their universality class. It is generally accepted that the effective
 low-energy theory for the  2D  Heisenberg systems with a collinear (N\'{e}el)
 order parameter is the $O(3)$
 non-linear sigma model (NLSM) in 2+1 dimensions \cite{nlsm}.
 This field theory contains a single effective coupling constant $g$
 and, at $T=0$, describes the ordered N\'{e}el phase for $g<g_{c}$.
 For $g>g_{c}$ the NLSM is in a quantum disordered phase with a finite
 correlation length. However the determination  of $g_{c}$ and the nature
 of the disordered phase are beyond  the field theory
 formulation and depend on the specific details of the model.
  In addition,  Berry phases associated with instanton tunneling
 between topologically different configuration are present in the
 NLSM \cite{berry}. In one dimension the Berry phase effects are
 known to be important, essentially leading to the difference between
 the excitations in the integer and half odd-integer spin chains
 \cite{affleck}. In 2D  Berry phases are also present but their role is
 less clear. If one  neglects these purely quantum effects,
 the universality class of the quantum transitions in the 2D HAFM
 should be the same as that of the classical $O(3)$ vector model
 in 3D \cite{exponents}. Quantum Monte Carlo simulations
 performed on the two-layer HAFM \cite{anders1} and the
 $CaV_4O_9$ lattice HAFM \cite{troyer} confirm with high accuracy
 that the quantum transitions in the above two models are in
 the $O(3)$ universality class. There also has been a report \cite{anders}
 that the 2D dimerized HAFM exhibits a deviation from the $O(3)$ behavior,
 which is presumably due to the small size lattices studied in the above work.
 Generally the most accurate Monte Carlo results seem to indicate
 that the quantum Berry phase effects are not important, at least
 in the models where the quantum transitions are driven by explicit
 (exchange driven) dimerization.
 
 On the other hand the $J_1-J_2$ model, which exhibits a quantum
 transition due to frustration, has a much better chance for deviation
 from the   $O(3)$ universality class.  
 The reason  is that the Berry phases  were shown to be relevant
 and intimately related to
 the presence of {\it spontaneous}  dimer (spin-Peierls) order
 in the quantum disordered phase of this model \cite{Read,subir}.
  Within the formalism of the large $N$ expansion for the $Sp(N)$ theory
 ($N=1$ being the physical limit), 
 Read and Sachdev  \cite{Read}  found two divergent length scales at the
 transition from the quantum disordered into the N\'{e}el ordered phase.   
 The first one is the usual correlation length $\xi$ which governs
 the exponential decay of the spin-spin correlations in the disordered phase
 and is  inversely proportional to the (triplet) magnon gap, 
 $\xi \propto 1/\Delta$. The instanton effects however lead to the 
 appearance of 
spontaneous spin-Peierls correlations and a second gapped singlet mode with a 
characteristic scale $\xi_{SP}$ (inverse singlet mass).
 The two scales are related via:

\begin{equation}
\label{sp}
\xi_{SP} \sim \xi^{C N}, \ \ N \gg 1, 
\end{equation}
 where $C = C_{1} + O(1/N), C_1 \sim 1$. Since $N$ is large
 one expects $\xi_{SP} \gg \xi$.  The presence of
 two divergent length scales at the transition point would  naively suggest
 a change in the universality class. However, as argued in Ref.\cite{subir1},
  the fact that $\xi_{SP}$ is a (large) power of $\xi$, causing the two
 scales to be well separated near the critical point, is a characteristic
 feature of a dangerously irrelevant coupling. This means that even though
 the Berry phases are relevant in the disordered phase, ultimately,
 near the critical point, their  effect disappears.
 In particular the dimer order parameter $D$ is expected to behave as
 $D \sim \xi_{SP}^{-1} \sim \Delta^{C N}$ and thus vanishes very fast
 as the critical point ($\Delta \rightarrow 0$)
 is approached \cite{subir}. In this scenario the quantum critical 
 fluctuations of the N\'{e}el order parameter are decoupled
 from the singlet mode and consequently the transition is still
 of $O(3)$ type. Notice that  the above analysis is certainly
 valid {\it provided} the $1/N$ expansion behaves well, since only then
 the $N \rightarrow \infty$ results  are relevant to  the physical
 situation $N=1$. However corrections beyond the  $N = \infty$ limit
 have not been systematically 
calculated in the literature, due to the complex nature of the
 problem.
 
 The purpose of the present work is to analyze the structure
 of the excitation spectrum  and the scales that appear
 near the quantum critical point in order to test the $Sp(N)$
 field theory predictions. We work directly with the physical
 spin problem ($N=1$) and the 
  approximation scheme that we use
 is based on  a perturbative expansion around the 
 spontaneously dimerized ground state in the quantum disordered phase.
 First, let us mention that the numerical implementation
 of this expansion via the dimer series expansion \cite{series,us1,plaq},
 as well as the  mean-field \cite{sub} and diagrammatic treatments \cite{us1},
 confirm the stability of the spontaneously dimerized phase  for
 intermediate values of frustration. This means that the large $N$ limit
 captures the essential physics of the problem, even though
 it can not be trusted
 numerically in regards to the exact location of the phase boundaries.
 However both the series expansions and the diagrammatic method are
 not accurate enough to calculate reliably the  critical exponents near the
 the transition into the N\'{e}el phase, since the exponents are not
 expected to vary considerably. 
  For example  the exponent $\nu$
 governing the vanishing of the triplet gap:  $\Delta 
\sim (g - g_c)^{\nu}$, is  $\nu \approx 0.71$ for
 the $O(3)$ and  $\nu \approx 0.75$ for the $O(4)$ universality
 class \cite{exponents}. Such small difference can not be confidently
 resolved with the above methods \cite{remark}. 
This is why we will follow a different route,
 namely  we will analyze the possibility of having additional soft modes
 at the transition (in addition to the triplet mode).


 The Hamiltonian of the  frustrated Heisenberg antiferromagnet is: 

\begin{equation}
\label{ham}
H = J_{1} \sum_{NN}{\bf S}_{i}.{\bf S}_{j} + J_{2} \sum_{NNN} {\bf
  S}_{i}.{\bf S}_{j},
\end{equation}
where $J_{1}>0$ is the nearest-neighbor, and $J_{2}>0$ is the frustrating
  next-nearest-neighbor  exchange coupling   on a square lattice (defined
 as shown in Fig.1).  
  All the spins $S_{i}=1/2$.  
In order to find  the excitation spectrum of $H$ we
 follow closely the treatment of Ref.\cite{us1} which is briefly
 outlined below.
 The starting point is grouping the spins into dimers (singlets) in the pattern, shown in Fig.1. This configuration (which is degenerate with 
  three others, obtained by translation by one lattice site,
  rotation by $\pi/2$, and rotation plus translation) was found
 to be stable in the parameter window $(J_{2}/J_{1})_{c1}
 < J_{2}/J_{1} < (J_{2}/J_{1})_{c2}$ \cite{us1}.
 Here  $(J_{2}/J_{1})_{c1} \approx 0.38$
is the transition point into the N\'{e}el phase, the neighborhood of
 which is the region we want to analyze.  
 The Hamiltonian can be expressed in terms of 
   bosonic  operators  $t_{i
  \alpha}^{\dag}, \alpha=x,y,z$ creating three degenerate triplet
 excitations from the singlets formed by each pair
  of spins, as shown in Fig.1. The site index $i$ now numbers the sites
 on the dimerized lattice.  
 The effective Hamiltonian describing the interactions between the
 triplets is \cite{us1}:

\begin{equation}
\label{ham1}
   H = H_{2} + H_{3} + H_{4}, 
\end{equation}
\begin{equation}
\label{ham2}
H_{2}  = \sum_{\bf{k}, \alpha} \left\{ A_{\bf{k}}
t_{\bf{k}\alpha}^{\dagger}t_{\bf{k}\alpha} +
\frac{B_{\bf{k}}}{2}\left(t_{\bf{k}\alpha}^{\dagger}
t_{\bf{-k}\alpha}^{\dagger} + \mbox{h.c.}\right) \right \}, 
\end{equation}
\begin{equation}
\label{ham3}
H_{3} =  \sum_{1+2=3} \mbox{R}({\bf k_{1}},{\bf k_{2}})
\epsilon_{\alpha\beta\gamma} t_{\bf{k_{1}}\alpha}^{\dagger}
t_{\bf{k_{2}}\beta}^{\dagger} t_{\bf{k_{3}}\gamma} + \mbox{h.c.}, 
\end{equation}
\begin{eqnarray}
\label{ham4}
H_{4} & = &  \sum_{1+2=3+4}  \left [ \mbox{T}({\bf k_{1}}-{\bf k_{3}})
(\delta_{\alpha\delta}\delta_{\beta\gamma}-
\delta_{\alpha\beta}\delta_{\gamma\delta}) \ + \right.
\\
&& \left.  \ \ \ \ \ \ \ \ \ \ \ \  
+ \  U \delta_{\alpha\delta}\delta_{\beta\gamma}
 \right ] t_{\bf{k_{1}}\alpha}^{\dagger}
t_{\bf{k_{2}}\beta}^{\dagger}t_{\bf{k_{3}}\gamma}
t_{\bf{k_{4}}\delta}. \nonumber
\end{eqnarray}
The  following definitions are used in Eqs.(\ref{ham2}-\ref{ham4}):
$
A_{\bf{k}}  =  J_{1} - (J_{1}/2)\xi_{k_{x}} + (J_{1} -
J_{2})\xi_{k_{y}} - J_{2}\xi_{k_{x}}\xi_{k_{y}}, \  B_{\bf{k}} =
 A_{\bf{k}} -  J_{1} 
$,
 and the matrix
elements in the quartic and cubic interaction terms:
$4\mbox{T}({\bf k})=J_{1} \xi_{k_{x}} +
 2(J_{1}+J_{2})\xi_{k_{y}} +
 2J_{2}\xi_{k_{x}}\xi_{k_{y}}, \ 
4 \mbox{R}({\bf
 p},{\bf q})= - J_{1} \gamma_{p_{x}} - 2 J_{2}
 \gamma_{p_{x}}\xi_{p_{y}} - \{p \rightarrow q\}
$, where we have defined  $\xi_{k} = \cos(k), \ \ \gamma_{k} = \sin(k)$.
 The $\mbox{T}$ and \mbox{R} terms describe the inter-site
 interactions arising from the exchange between the dimers.
 An additional on-site ($U$) term is also introduced 
 and one must set $U \rightarrow \infty$.
 This term reflects the hard-core nature of the bosons which
 follows from 
  the kinematic constraint on the Hilbert space $t_{i
 \alpha}^{\dag} t_{i \beta}^{\dag} = 0$.
 The constraint is necessary in order to ensure that the bosonic
 Hamiltonian in terms of the triplet operators corresponds uniquely to
 the original spin Hamiltonian (\ref{ham}) and no unphysical states
 appear in the final result. The sums over $\bf{k}$ extend over the
 Brillouin zone of the dimerized lattice, i.e. $-\pi \leq k_{x}, k_{y}
 \leq \pi$. In this notation the N\'{e}el ordering wave-vector
 ($(\pi,\pi)$ of the original lattice) is ${\bf Q}_{AF} = (0,\pi)$.

 The spectrum of Eq.(\ref{ham1}) was studied in Ref.\cite{us1} by 
summing selected
 infinite series in the perturbative expansion. The dilute Bose gas
 approximation was used and the diagrams  classified in powers of
 the density of magnons. 
The diagrammatic treatment
 was also compared with numerical results obtained by high-order dimer
 series expansions and the agreement was found to be very good.
 We will therefore present only diagrammatic results from now on.  
In the quantum disordered
 phase  $(J_{2}/J_{1})_{c1}
 < J_{2}/J_{1} < (J_{2}/J_{1})_{c2}$ the triplet excitation
 spectrum $\omega({\bf k})$ has a non-zero gap
 $\Delta =\omega({\bf Q}_{AF})$ which reflects the fact that the dimer
 configuration is stable. As the critical point  $(J_{2}/J_{1})_{c1}=0.38$
 is approached,  $\Delta \rightarrow 0$,   signaling an instability towards 
 a phase with  non-zero  N\'{e}el order parameter. 
 The variation of $\Delta$ as a function of frustration is shown in
 Fig.2. Let us mention  that as we move close to the critical point from the
 disordered side,  the density of triplets increases and
  is approximately $0.3$ at $J_{2}/J_{1} = 0.38$. This leads, 
 in principle, to a $30 \%$ uncertainty in the results due to the
 omitted higher order diagrams. 
However the  accuracy in the position of the critical point (0.38) is much
better than 30\% because of the steep dependence of the gap on $J_2/J_1$.
Nevertheless, within the accuracy of our calculation  the point where 
  $\Delta \approx 0.05 J_{1}$ (see Fig. 2) is practically indistinguishable
from  the critical point.


 In Ref.\cite{us1} it was pointed out that an additional collective low-energy
 mode also exists near   $(J_{2}/J_{1})_{c1}$. This excitation is
 a bound state of two triplets with total spin $S=0$.
 We proceed to investigate its properties in more detail.
 Introducing the total (${\bf Q}$) and the relative (${\bf q}$)
 momenta of the two triplets forming the bound state, the two-particle
 singlet  is:

\begin{equation}
|\Psi_{{\bf Q}}\rangle= \sum_{{\bf q}, \alpha}\Psi
({\bf q},{\bf Q})t_{\alpha, {\bf Q}/2 + {\bf q}}^{\dagger} 
t_{\alpha, {\bf Q}/2 - {\bf q}}^{\dagger}|0\rangle.
\end{equation}
The bound state wave-function $\Psi({\bf q},{\bf Q})$ satisfies the
integral equation:
\begin{eqnarray}
\label{BS}
\lefteqn{
\left [E^{S}({\bf Q})-\omega_{{\bf Q/2}+{\bf q}}-\omega_{{\bf Q/2}- {\bf q}}
\right ]\Psi({\bf q},{\bf Q})=}  \nonumber \\
& & 
 \int\frac{d{\bf p}}{(2\pi)^{2}} \left\{ -2[\mbox{T}({\bf p}-{\bf q}) +
 \mbox{T}({\bf p}+{\bf q})] + U \right \}
\Psi({\bf p}, {\bf Q}),
\end{eqnarray}
 which can be easily derived by noticing that it is equivalent
 to the two-particle Schr\"{o}dinger equation: $H|\Psi_{{\bf Q}}\rangle = 
 E^{S}({\bf Q}) |\Psi_{{\bf Q}}\rangle$. Here $E^{S}({\bf Q})$ is the energy
 of the collective mode.
 The function $\mbox{T}({\bf q})$ is the two-particle scattering
 amplitude from (\ref{ham4}):

\begin{equation}
\label{T}
\mbox{T}({\bf k})= \frac{J_{1}}{4} \cos{k_{x}} + 
\frac{(J_{1}+J_{2})}{2}\cos{k_{y}} +
 \frac{J_{2}}{2}\cos{k_{x}}\cos{k_{y}}.
\end{equation}
 This interaction leads to attraction between two triplets in the 
 singlet channel.
 In Eq.(\ref{BS}) the (second-order) contribution of
 $H_{3}$, Eq.(\ref{ham3}) into binding has been neglected. We have checked
 that the perturbative inclusion of this term indeed  leads only to a small change
 of the results presented below. 
Since  we have to take $U \rightarrow \infty$,
 the following replacement must be made on the right hand side of
 Eq.(\ref{BS}):
  $U\int d{\bf p}\Psi({\bf p}, {\bf Q}) \rightarrow \lambda$, where
 $\lambda$ is a Lagrange multiplier,
 determined self-consistently from the condition  
$\int d{\bf q} \Psi({\bf q}, {\bf Q}) = 0$.
 The bound state exists only if a solution of Eq.(\ref{BS}) can be found 
 such that $E^{S}({\bf Q}) < E_{c}({\bf Q}) = \mbox{min}_{{\bf q}}[\omega({\bf Q
}/2 +{\bf q})
 + \omega({\bf Q}/2 -{\bf q})]$, meaning that it must be below the
 two-particle scattering continuum $E_{c}$.
 In  Fig.3 we present the  numerical solution of Eq.(\ref{BS})
 for a fixed value of frustration ($J_{2}/J_{1}=0.4$)  above the critical value.  
 We  have also plotted the one-particle spectrum $\omega({\bf k})$, and
 the  shaded region is the two-particle scattering region $E>E_c({\bf k})$.
 The bound state is stable (non-zero gap)  
for all $\bf k$ throughout the disordered phase, with a minimum of the
 dispersion  $E^{S}({\bf k})$ at ${\bf k} = (0,0)$. As frustration
 decreases 
  the singlet gap  $E^{S}(0,0)$ decreases 
 and  appears to vanish at the critical point, as shown in Fig.2. 
 
 We believe that the existence of a singlet
 bound state at ${\bf k} = (0,0)$ reflects the
 {\it spontaneous} nature of the dimer order. Indeed,  we have checked
 that in models where the dimerization is explicit, i.e. due
 to stronger exchange on certain bonds, 
  the singlet does not exist in the neighborhood
 of ${\bf k} = (0,0)$, meaning that its binding energy is zero.
  The binding energy is defined as $\epsilon({\bf k}) = E_c({\bf k}) 
 - E^{S}({\bf k})$. We have found that $\epsilon(0,0) =0$ 
 both in the 2D dimer model (as defined in Ref.\cite{dimer})  and in the
 two-layer model (Ref.\cite{chub}). 
Unlike the above two models, the Hamiltonian Eq.(\ref{ham}) does not 
 break any lattice symmetries, but the ground state of Fig.1 certainly does.
It was argued in Ref.\cite{us1} that the vanishing of the
spontaneous dimer order at the critical point is intrinsically related to
the low energy singlet mode.
Thus we believe that the singlet bound state 
reflects the presence  of   
   non-zero dimer order parameter   in the $J_1-J_2$ model, 
 i.e. the spontaneous breakdown of the discrete lattice symmetries
 in the quantum disordered phase.

  Even though we have shown that the gaps for both the triplet and singlet
 modes vanish at the critical point, this does not necessarily
 mean a departure from the $O(3)$ universality class which
is  related to the triplet only.
In what follows we will in fact demonstrate that the singlet mode is 
 "irrelevant" at 
the critical point.
 Let us start with the observation that
the relevancy (or otherwise) of a soft excitation is directly related
 to its spectral weight. We find, as expected, that the spectral weight
 of the  triplets (the residue of the one-particle
 Green's function)  stays finite at the transition point.
 However, the spectral weight of the collective singlet excitation 
  is proportional to the binding energy (which in turn is inversely
 proportional to the size of the bound-state wave function).
 Observe that the lower edge of the
 two-particle continuum at ${\bf k}=(0,0)$, 
 $E_c(0,0) = 2 \omega({\bf Q}_{AF}) = 2 \Delta$,
 and since at the transition point $\Delta \rightarrow 0$,
 the binding energy must vanish as well, $\epsilon(0,0) \rightarrow 0$.
 The variation of the binding energy as a function of frustration,
 obtained by solving Eq.(\ref{BS}) numerically,
 is shown in Fig.2. Since we realize that for $\Delta \rightarrow 0$
 the accuracy of the calculation decreases, let us find the asymptotic
 behavior of $\epsilon(0,0)$ in this limit  analytically.
 The one-particle dispersion around its minimum 
 has the form:
 $\omega^{2}({\bf k}) =  \Delta^{2} + c^{2}|{\bf k}-{\bf Q}_{AF}|^{2}$,
 which is valid close to the critical point. The triplet
 velocity $c$ is known to remain finite at the transition 
 \cite{linear,igarashi,exact1,ising}, 
 $\Lambda c \sim J_1$, where $\Lambda \sim 1$ (in units of the inverse
 lattice spacing) is a characteristic momentum. 
 Denoting the right hand side of Eq.(\ref{BS}) by $\Phi({\bf q},{\bf Q})$,
 the solution of  (\ref{BS}) at ${\bf Q}=(0,0)$ is:
 $\Psi({\bf q},0) = \Phi({\bf q},0)[E^{S} - 2 \omega({\bf q})]^{-1}$.
 We define $E^{S} = E^{S}(0,0)$ and  $\epsilon = \epsilon(0,0)$ from now on. 
 The energy can be found from the  condition:
 $\int d{\bf q} \Psi({\bf q},0 ) = 0$. This integral diverges
 logarithmically at ${\bf Q}_{AF}$ for small binding:

\begin{equation}
\label{int}
\int \frac{ d^{2}q}{2 \omega({\bf q})
- E^{S}} \sim \frac{\Lambda}{c}
 +\frac{ 
\Delta}{c^{2}} \ln{\left (\frac{\Lambda c}{\epsilon} \right )},
 \ \ \epsilon \ll \Delta \ll J_1.
\end{equation} 
We remind that $\epsilon = 2\Delta -E^{S}$. 
 When estimating the integral,  $\Phi({\bf q},0)$
 can be replaced by $\Phi({\bf Q}_{AF},0)$.
 This quantity is finite at the critical point which follows
 from Eq.(\ref{T}). 
From (\ref{int}) we find the binding energy:

\begin{equation}
\label{binding}
\epsilon = \epsilon_0 \exp \left ( {-\frac{\epsilon_1}{\Delta}} \right ),
\ \  \Delta \ll J_1, 
\end{equation}
where $\epsilon_0,\epsilon_1 \sim J_1$ are two constants that depend
 weakly on $J_2$ and are finite at the critical point.
 The result is similar to the  formula for the exponentially
 small s-wave bound state in a 2D potential \cite{landau}.
 However there are  two differences from the usual expression \cite{landau}:
1.) the pre-exponential factor in Eq.(\ref{binding})
does not depend on the "mass" $\Delta$ because of the relativistic form of the
 dispersion near the transition point, and  2.)
because of the hard core constraint ($U \rightarrow \infty$) 
 the exponent $\epsilon_1$ can not be
written as $|\int {\cal{U}} dr|^{-1}$, where ${\cal{U}}$ is the attractive
 potential (i.e. the expression in the curly brackets in Eq.(\ref{BS})). 

In Fig.2 we present a fit of the formula 
(\ref{binding}) to the numerical solution of (\ref{BS}).
The major disagreement occurs only close to the critical point,
 for $J_{2}/J_{1} < 0.41$.
 The asymptotically exact solution predicts an exponentially
 fast vanishing of $\epsilon$ which can not be captured
 reliably in the numerical solution.

The size $R$ of the bound state (in units of the lattice spacing)
 is determined by the spatial
extent of the wave-function: 
$R^{2}= \sum_{\bf q} |\partial \Psi({\bf q})/ \partial {\bf q}|^{2}$,
 where $\Psi$ is assumed normalized. Evaluating this expression 
 for $\Delta \rightarrow 0$ leads to

\begin{equation}
R^{2} \sim \frac{J_{1}^{2}}{\epsilon \Delta} \sim \xi \exp{(C_{\xi}\xi)},
 \ \ \xi \gg 1,
\end{equation}
 where $\xi \sim \Delta^{-1}$ is the correlation length, and
$C_{\xi} \sim 1$ is a constant.  Thus we find, as expected, that
 the size increases as the binding energy decreases. At the critical point
 $R$ diverges exponentially  with the correlation length.
 This in turn implies that the spectral weight of the collective singlet 
 vanishes exponentially fast. Consequently the singlet bound state
 does not influence the
triplet dynamics near the critical point
 and hence can not change the $O(3)$ universality class.

Unlike  the theory of Read and Sachdev \cite{subir},  our approach
 does not relate directly 
 the gap in the singlet spectrum and the 
 dimer order parameter. The  latter quantity is defined as
 $D = \langle {\bf S}_{2}.{\bf S}_{3} \rangle - 
  \langle {\bf S}_{1}.{\bf S}_{2} \rangle$ (see Fig.1.).
 We  have presented
 arguments  \cite{us1} that the low-energy singlet
 affects the dimer order by increasing the quantum fluctuations.
 This effect becomes stronger and stronger as the critical
 point is approached and one could ultimately expect that $D$ vanishes.
 High order dimer series results support this conclusion \cite{us1}, but
 can not determine  the critical behavior of $D$.
 Thus we can not verify the prediction of the large-$N$ theory 
 (which follows from Eq.(\ref{sp})) that
 $D$ should vanish with a large exponent.    

In summary, we have found that  
 in the 2D $J_1-J_2$ model the critical behavior near the transition
 between the spontaneously dimerized and the N\'{e}eel phase is
 characterized by two soft modes - the usual triplet magnon mode,
 and a collective singlet excitation. Even though the gap 
in the singlet spectrum  vanishes at the transition, we argue that it does not
influence the critical dynamics of triplet excitations.
The reason  is that
 the spectral weight of the singlet vanishes exponentially fast 
 at the quantum critical point.
We have to note that our picture is different from that of Read and
Sachdev \cite{Read,subir}, based on the large-$N$ expansion. 
 In \cite{Read,subir}  the singlet gap was found to be  much smaller
than the triplet one, which in turn leads to the second large
 length scale. In our picture the
 singlet gap is approximately equal to two triplet gaps, and the large 
length scale comes from the size of
the singlet bound sate. Nevertheless the final conclusion concerning
the triplet critical dynamics is the same: 
 the $O(3)$ universality class describes the transition between 
the  N\'{e}el ordered and quantum disordered phase. 

Finally let us mention that in cases where the ordered
 phase is characterized by incommensurate correlations 
 (e.g. the triangular lattice Heisenberg model) the critical behavior
 could be quite different. On semi-classical level the order
 parameter has $SO(3)$ symmetry and consequently an $O(4)$
 universality class is possible \cite{azaria}.
 If indeed the $O(4)$ class is realized  in this case,  the singlet mode
 should become a truly Goldstone mode at the transition and
 therefore must have very different properties from the ones
 found in the present work.

\ \

It is our pleasure to thank Subir Sachdev, N. Read,  
 Cristiano Biagini, and Sergei Obukhov  for  many stimulating discussions.
  V.N.K.  acknowledges the financial  support of
 NSF Grant DMR9357474.

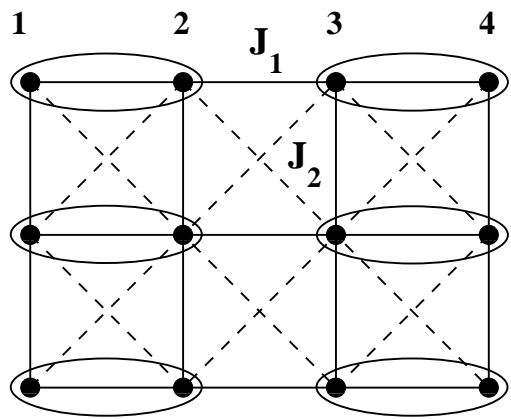
\begin{figure}
\caption
{The $J_1-J_2$ model on a square lattice.
The circles represent spins paired in singlets in the columnar ladder
 dimerization pattern.
}
\label{fig.1}
\end{figure}

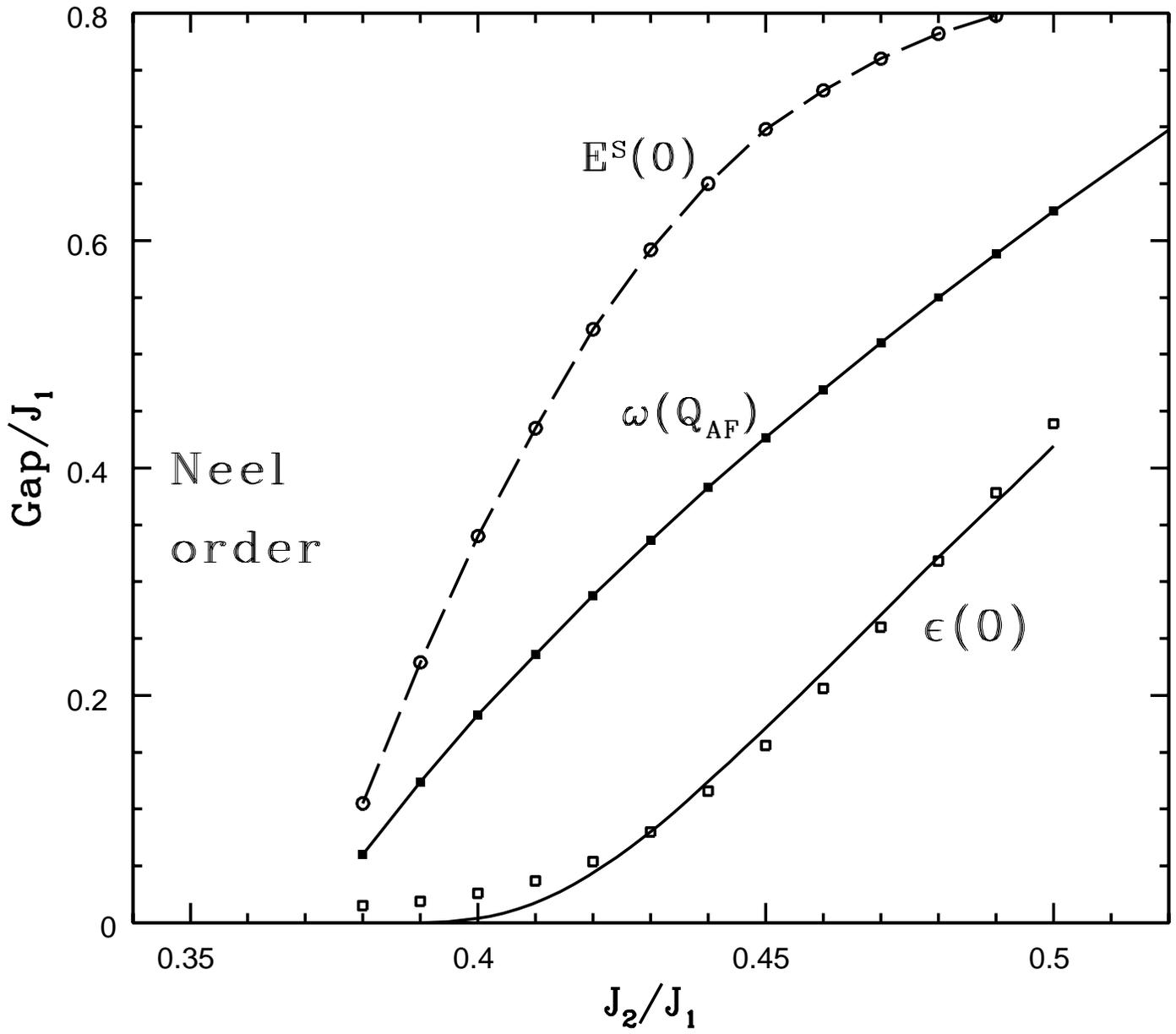
\begin{figure}
\caption
{Gaps in the one and two-particle excitation spectra.
 Solid squares, connected by a solid line, represent the
 triplet gap $\Delta=\omega({\bf Q}_{AF})$, and open circles
 connected with a dashed line represent the gap of the two-particle
 singlet bound state at ${\bf k}=(0,0)$. 
 Open squares  are the singlet binding energy, obtained by
 solving Eq.(\ref{BS}) numerically, while the solid line is 
 the fit, based on the asymptotic formula,  Eq.(\ref{binding}),
 with $\epsilon_0=2.85J_{1},\epsilon_1=1.20J_{1}$.
 }
\label{fig.2}
\end{figure}

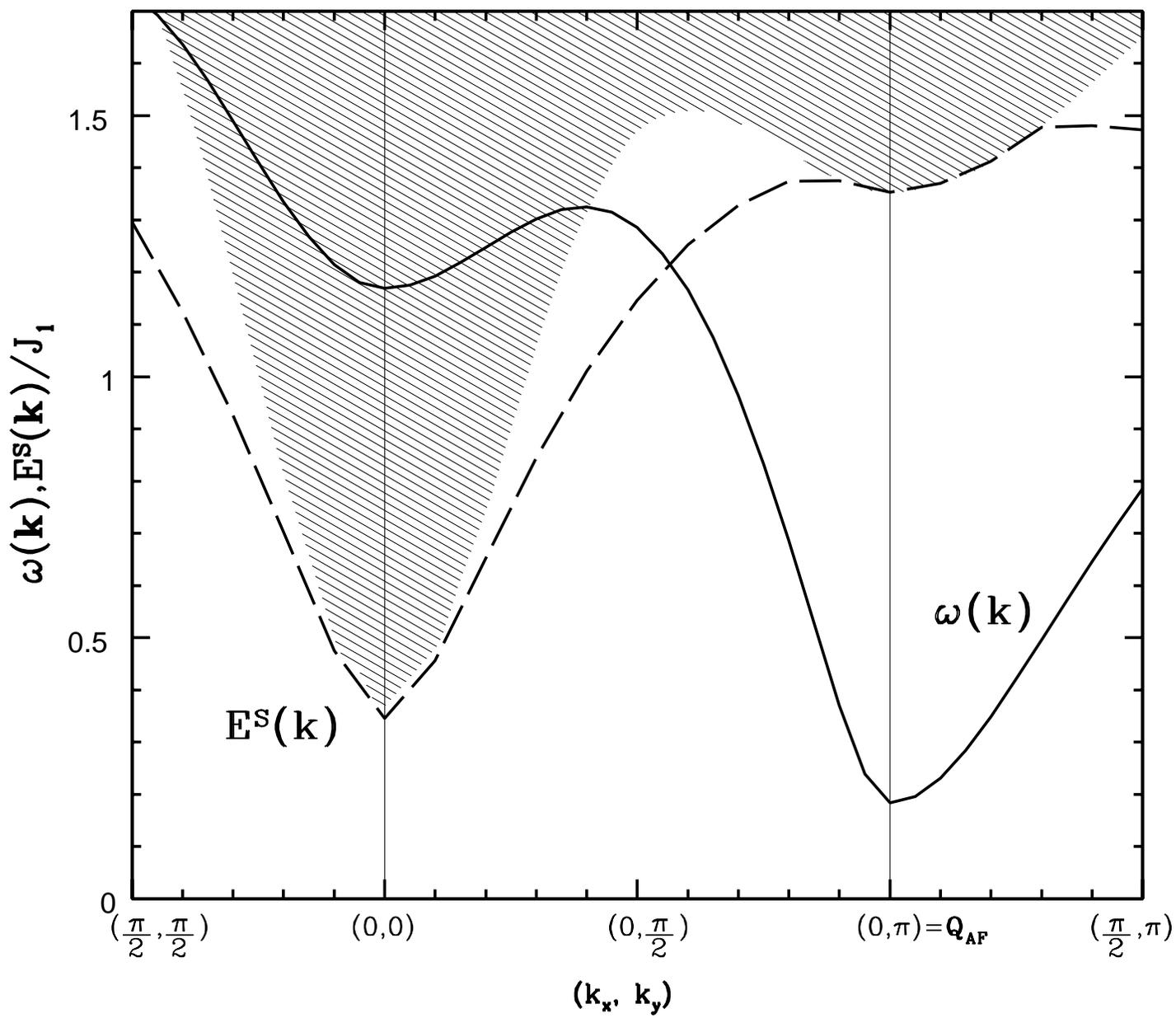
\begin{figure}
\caption
{ Triplet magnon excitation spectrum (solid line), and
 the singlet bound state excitation branch (long-dashed line)
 in the part of the Brillouin zone near $(0,0)$ and ${\bf Q}_{AF}$.
 The shaded area represents the two-magnon scattering continuum.
 All spectra are calculated diagrammatically for $J_{2}/J_{1}=0.40$.
 }
\label{fig.3}

\end{figure}


\begin{references}
\bibitem{remark}  Unfortunately the potentially most powerful
 tool for calculating exponents - the quantum Monte Carlo method, can
 not be applied to models with frustration since it suffers from the
 notorious sign problem.
\end{references}
\end{document}